\documentclass{sf2a-conf2018}
\usepackage{natbib}
\usepackage{diagbox}
\usepackage{slashbox}
\usepackage{multirow}
\usepackage{float}
\usepackage{booktabs}
\usepackage{graphicx,subfigure}
\usepackage{txfonts}
\usepackage[]{hyperref}
\usepackage{appendix}
\usepackage{epstopdf}
\usepackage{color}

\def\BibTeX{{\rm B\kern-.05em{\sc i\kern-.025em b}\kern-.08em
    T\kern-.1667em\lower.7ex\hbox{E}\kern-.125emX}}
\bibpunct{(}{)}{;}{a}{}{,}  

\newcommand{\G}{\mathcal{G}}

\begin{document}

\TitreGlobal{SF2A 2018}


\title{The limit of the gyrochronology - A new age determination technique: the tidal-chronology}


\runningtitle{}

\author{F. Gallet}\address{Univ. Grenoble Alpes, CNRS, IPAG, 38000 Grenoble, France}

\author{P. Delorme$^1$}







\setcounter{page}{237}


\maketitle


\begin{abstract}
While the number of detected planets is continuously increasing since 1995, their impact on their central star still remain poorly understood. Yet, the presence of a massive close-in planet can strongly modify the surface angular velocity evolution of the star. In these circumstances, age estimation techniques based on rotation, such as the gyrochronology, can't be applied to stars that experienced significant star-planet magnetic and tidal interactions during their evolution.  With the use of a numerical model that combines an angular momentum evolution and an orbital evolution code, we investigated the evolution of initial distribution of star-planet systems, in which the orbital evolution of the planet is driven by the tidal dissipation formalism. Based on these initial distributions we highlighted the limits of applicability of the gyrochronology analysis and proposed a new age determination technique based on the observation of the couple $\rm P_{rot,\star}$-$a$, where $\rm P_{rot,\star}$ is the stellar rotational period and $a$ the planetary semi-major axis. This technique called tidal-chronology can be very helpful for planetary system composed of a star between 0.3 and 1.2 $\rm M_{\odot}$ and a planet more massive than 1 $\rm M_{jup}$ initially located at few hundredth au of the host star.

\end{abstract}

\begin{keywords}
planet-star: interactions -- stars: evolution -- stars:rotation
\end{keywords}


\section{Introduction}

Estimating the age of a planetary system is fundamental to constrain: 1) current planetary models \citep{Ida08,Mordasini09,Mordasini12,Alibert13}; 2) star-planet interaction efficiency \citep{Lanza11}; 3) internal structure and chemical composition of the star (based on stellar model); and 4) the star-planet system previous evolution \citep{Gallet17}.

Empirical age determination techniques such as the gyrochronology \citep{Barnes03} and the magnetochronology \citep{Vidotto14} are proposed in the literature. Both are based on the observation that during the main-sequence phase (hereafter MS) the evolution of the surface rotation and magnetic field of a star, for a given stellar mass, only depend on age. 
However, as long as the star departs from an isolated state, i.e. if a massive planet orbits around the star, these techniques can no longer be used since star-planet tidal interaction could have modified the evolution of the surface rotation rate along the system's evolution \citep[see][]{Gallet18}. As a consequence, the age of several planetary systems hosting a hot Jupiter could appear younger than they really are.

We aim to provide the community with a new age estimation technique based on the modelling of the evolution of the star-planet tidal interaction and observation of the surface rotation rate of the host star and current location of the massive planet orbiting it.

\section{Numerical model}

\subsection{Description}


In this work we combined the stellar rotational evolution model described in \citet{GB15} to the modified orbital evolution model used in \citet[][]{Bolmont16}. The link between the two is done through the grid of tidal dissipation from \citet[][]{Gallet17b}. The stellar structure is provided by the stellar evolution code STAREVOL \citep[see][and references therein]{Amard15}.

The aim of this coupling is to follow, through a realistic approach, the evolution of a given star-planet system. This code is specifically designed for stars between 0.3 (fully convective limit) and 1.2 $\rm M_{\odot}$, and for planetary systems composed of only one massive planet.
\begin{figure*}[!ht]
    \begin{center}
    	\subfigure[$\rm M_{p} = 2~M_{jup}$]{
            \label{DeltaP3D_2Mjup}%
            \includegraphics[width=0.48\linewidth]{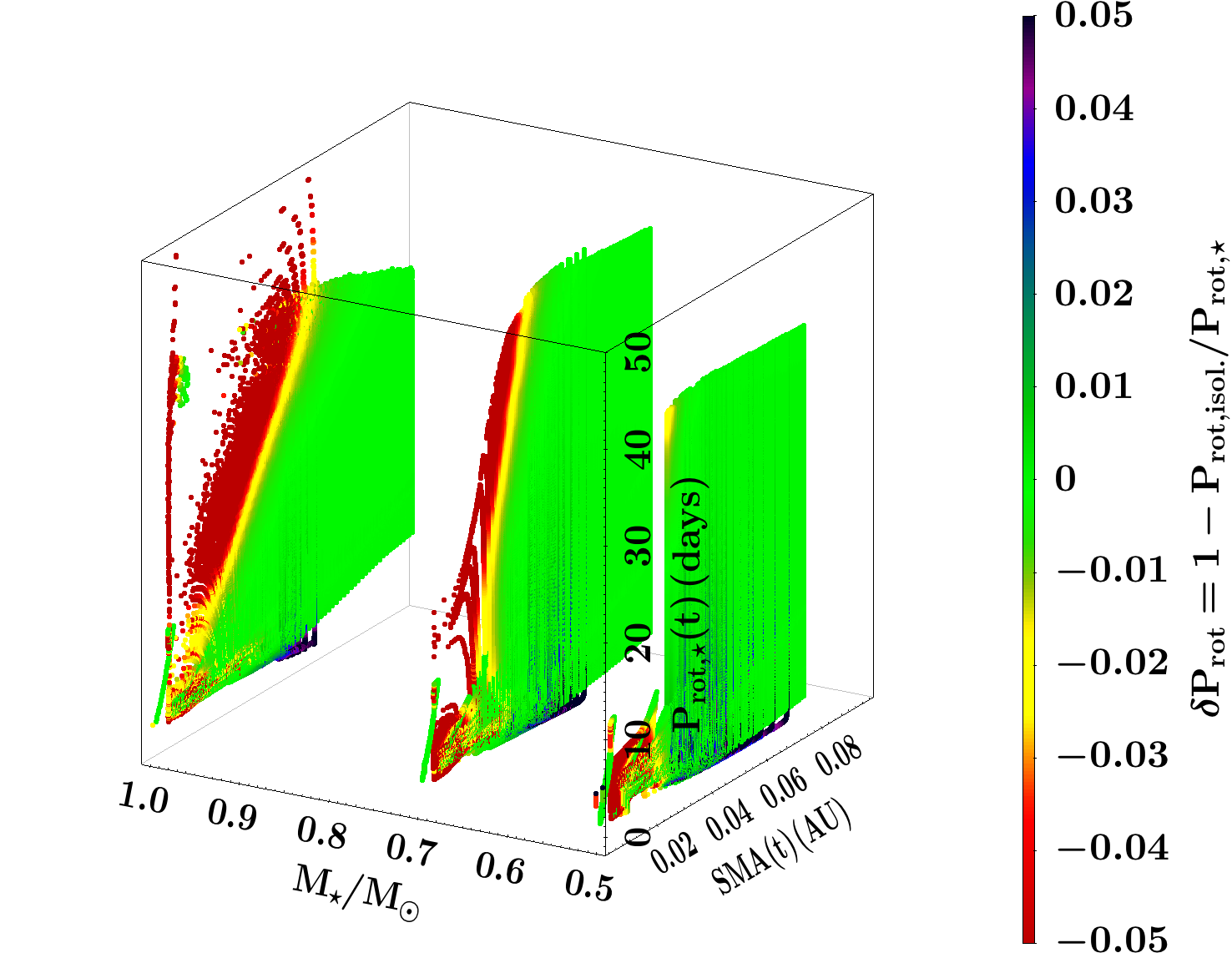}
        } 
       \hspace{-0.1cm}
        \subfigure[$\rm M_{p} = 5~M_{jup}$]{
           \label{DeltaP3D_5jup}%
           \includegraphics[width=0.48\linewidth]{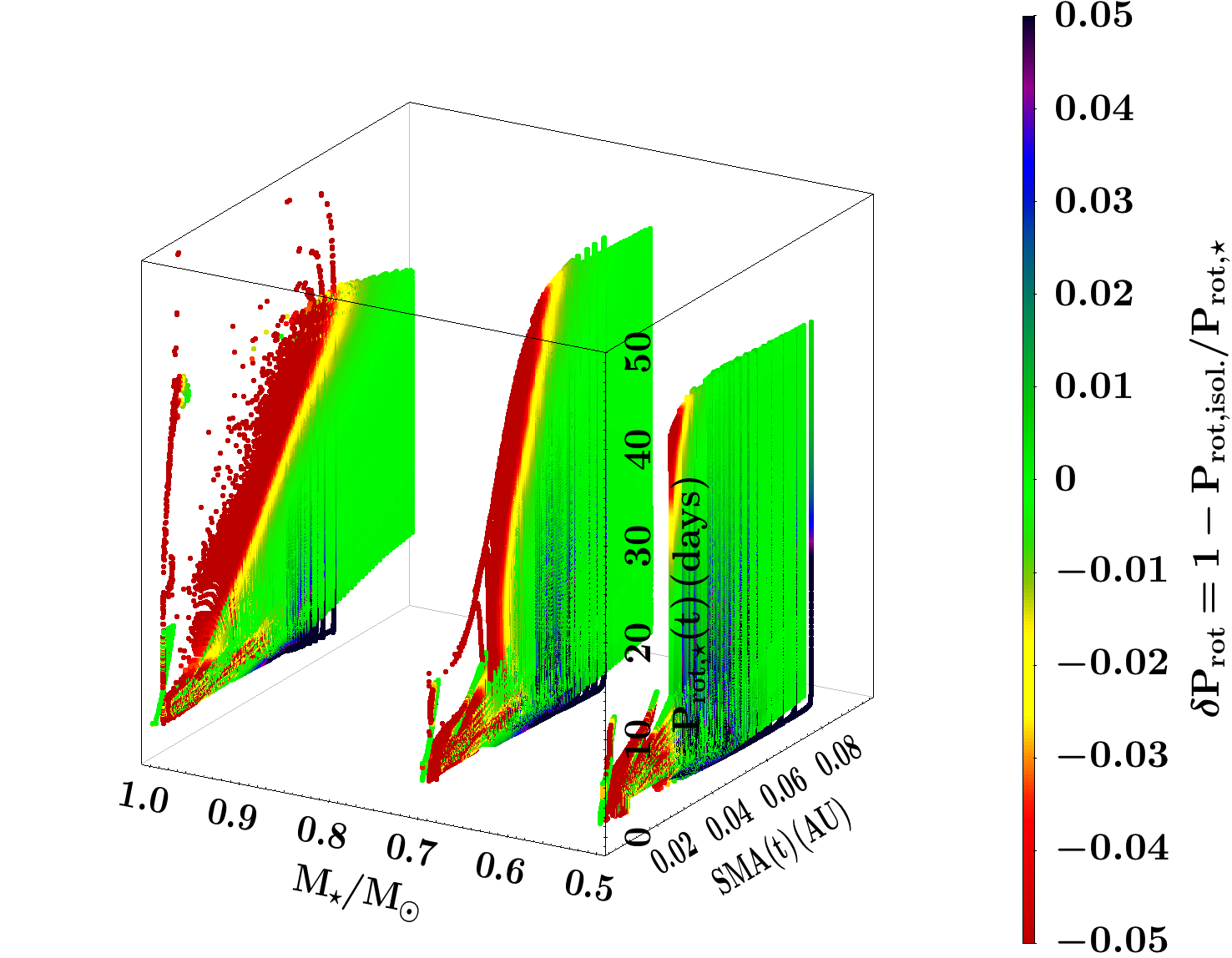}
        } 
        \caption{In green, region in which the gyrochronology analysis can be used. {$\rm P_{rot,\star} (t)$ as a function of $\rm SMA (t)$ and stellar mass for a 2 (left) and 5 (right) $\rm M_{jup}$ mass planet. The color gradient corresponds to the rotational departure $\rm \delta P_{rot}=1 - P_{rot,isol.}/P_{rot,\star}$. Only systems for which the planet is still present are plotted.}}
        \label{DeltaP3D}%
    \end{center}
\end{figure*}

Using this meta-code, we computed a grid composed of 0.5, 0.7, and 1.0 $\rm M_{\odot}$ with initial rotational period between one and 11 days \citep[using a parametrization that follows][]{GB15,Gallet18}. We considered a 2 $\rm M_{jup}$ and a 5 $\rm M_{jup}$ mass planet initially located between 0.1 and 1.0 $\rm R_{co}$, with the corotation radius $\rm \rm R_{\rm co} = \left( \displaystyle \frac{\G M_{\star} }{\Omega_{\star}^2}  \right)^{1/3} = \left( \displaystyle \frac{\G M_{\star} P_{\rm rot,\star}^2 }{(2\pi)^2}  \right)^{1/3},$ where $\rm P_{\rm rot,\star}$ is the surface rotation period of the host star, $\G$ the gravitational constant, $\rm M_{\star}$ the stellar mass, and $\Omega_{\star} = 2\pi / \rm P_{\rm rot,\star}$ the surface angular velocity of the host star.

\subsection{The gyrochronology and its domain of applicability}

The gyrochronology technique was initially proposed by \citet{Barnes03} as a way to estimate the age of isolated stars. It is based on the behaviour of the surface rotation rate of the stars (between 0.3 and 1.0 $\rm M_{\odot}$) that seems to evolves as $t^{-1/2}$ during the MS phase \citep[see][]{Sku72}. 

Figure \ref{DeltaP3D} shows the evolution of $\rm P_{rot,\star}$(t) as a function of  the semi-major axis (hereafter SMA), stellar mass (0.5, 0.7, and 1.0 $\rm M_{\odot}$), and planetary mass (2 $\rm M_{jup}$ and 5 $\rm M_{jup}$). The color gradient corresponds to the variation of $\delta \rm P_{rot}$ ($\pm 5\%$) for each couple $[\rm P_{rot,\star}(t)-SMA(t)-M_{\star}-M_p]$, where $\rm \delta P_{rot} = 1 - P_{rot,isol.}/P_{rot,\star}$ traces the departure of the rotational evolution of star in star-planet system to the rotation evolution of the same but isolated (i.e. without planet) star. Figure \ref{DeltaP3D} displays the domain of validity of the gyrochronology analysis (the green parts). In this figure the reddest parts correspond to the region where the star is 5\% faster ($|\delta \rm P_{rot}| \geq 5\%$) than an isolated star, which corresponds, following the Skumanich relationship \citep[$\rm \Omega_{\star} \propto t^{-1/2}$,][]{Sku72}, to an error of about 10\% on the age estimation using the gyrochronology analysis \citep{Delorme11}. Increasing the planetary mass slightly extends this region toward lower SMA(t).

If a massive close-in planet is detected orbiting its host star, then the gyrochronology analysis can't be applied if the planet is more massive than about 1 $\rm M_{jup}$ and located below 0.1 au. If no planets are detected but that there is a suspicion of planetary engulfment then the gyrochronology analysis can't be used. In some cases, the gyrochronology can't be applied even several Gyr after the engulfment (see Gallet \& Delorme in prep.) as the footprint of this interaction survive a large fraction of the stellar rotational evolution history.

\section{Tidal-chronology}

Following the work of \citet{Gallet18} we developed a new age estimation technique based on the measurement of both the surface rotation rate of the star and location of the planet around it. It relies on the fact that the star-planet interaction produces {a rotation cycle that can be used to estimate the age of a given close-in system. Since the rate of the evolution of the SMA strongly depends on its value \citep[see Eqs. 3-6 from][]{Gallet18}, a given observed couple $\rm P_{\rm rot,\star}(\rm t)-\rm{SMA(t)}$ is thus only produced at a small range of possible ages and initial conditions. 
\begin{figure*}[!ht]
    \begin{center}
    	\subfigure[$\rm M_{p} = 2~M_{jup}$]{
            \label{Degemass3D_2Mjup}%
            \includegraphics[width=0.48\linewidth]{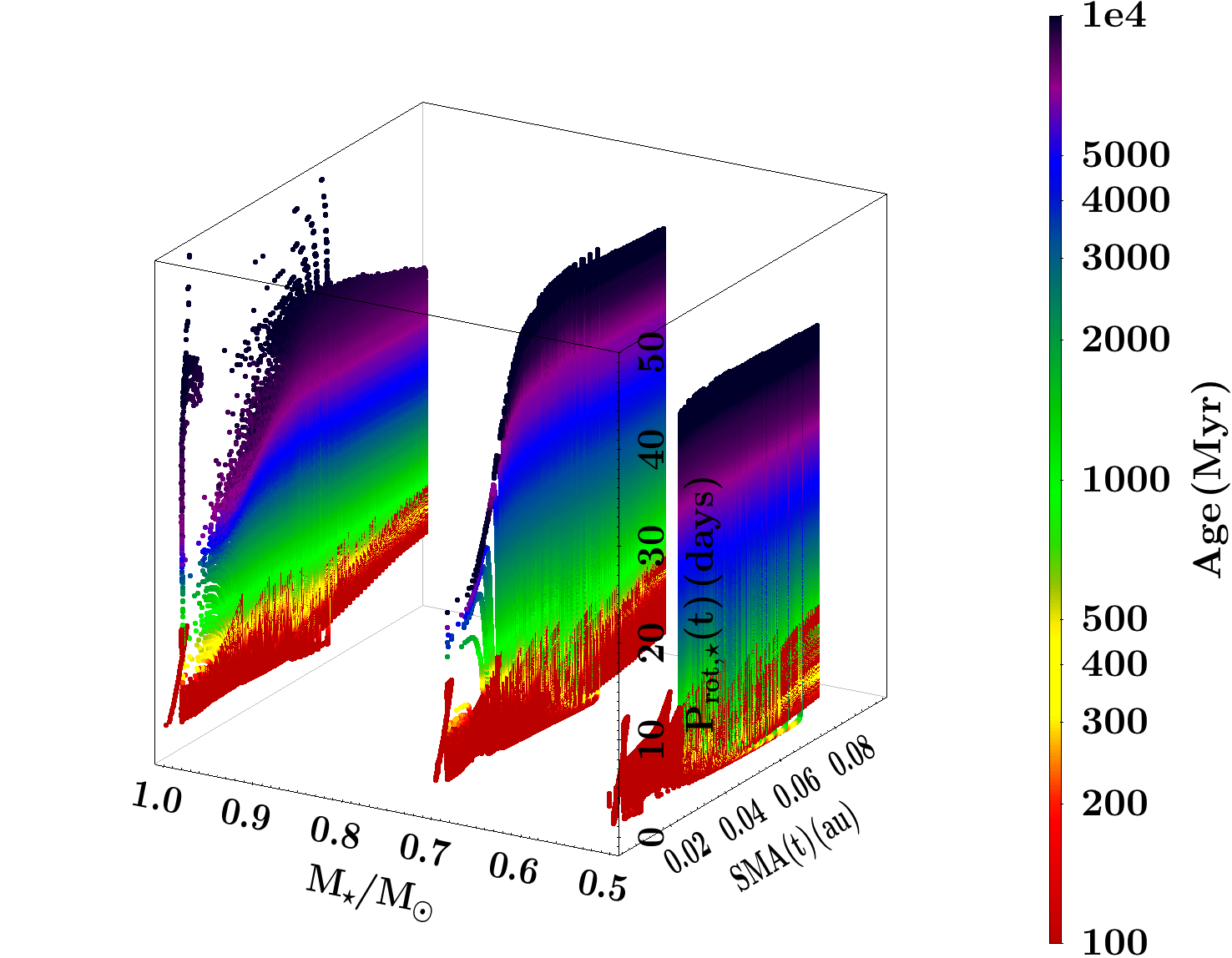}
        } 
       \hspace{-0.1cm}
        \subfigure[$\rm M_{p} = 5~M_{jup}$]{
           \label{DDegemass3D_5jup}%
           \includegraphics[width=0.48\linewidth]{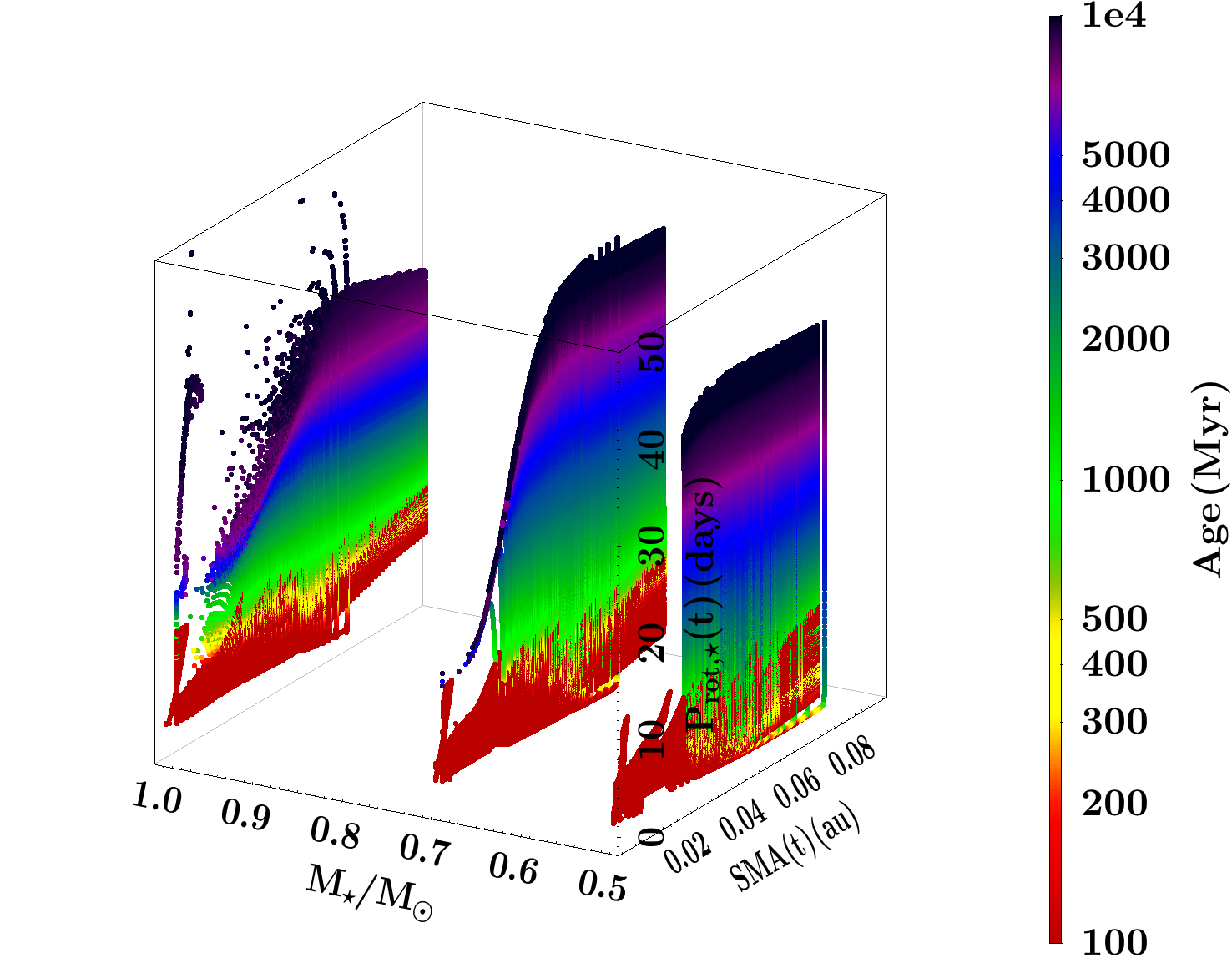}
        } 
        \caption{Synthetic $\rm P_{\rm rot,\star}$(t) and SMA(t) estimated for a system composed of a 0.5, 0.7, and 1.0 $\rm M_{\odot}$ star around which orbits a 2 and a 5 $\rm M_{\rm jup}$ planet. $\rm P_{rot,init}$ = 1-11 days (with $\Delta \rm P_{rot,init}$ = 0.2 days) and $\rm SMA_{init}=$  0.1, 0.3, 0.4, 0.45, 0.5, 0.55, 0.7, 0.8, 0.92, 0.94, 0.96, 0.98, and 1.0  $\rm R_{co}$ \citep[see ][]{Gallet18}. The color gradient depict the age (in Myr) at which the couple $\rm P_{\rm rot,\star}$(t)-SMA(t) is extracted. Only planetary systems in which the planet is still orbiting the star are plotted.}
\label{Degemass3D}%
    \end{center}
\end{figure*}


To determine the age of the system we computed a grid of star-planet system's evolution composed of stars with $\rm P_{\rm rot, init}$ between one and 11 days and planetary initial SMA between 0.1 and 1.0 $R_{\rm co}$ for a given stellar and planetary mass that match the observed system's properties. We then explore the grid so as to extract at which age the observed couple $\rm P_{\rm rot,obs}-\rm SMA_{obs}$ is retrieved. 

We finally estimate the departure of the observed couple $\rm P_{\rm rot,obs}-\rm SMA_{obs}$ to each of the $\rm P_{\rm rot,\star}$-SMA$_\star$ couples from the grid using this chosen expression:
$$
S^2 = \frac{  (\rm P_{\rm rot,\star}-\rm P_{\rm rot,obs})^2  }{\rm P_{\rm rot,obs}^2} + \frac{(\rm SMA_{\rm \star}-\rm SMA_{\rm obs})^2}{ \rm SMA_{\rm obs}^2}.
$$
We also applied a 3D interpolation method using the \textit{Python-SciPy} griddata routine \citep[to interpolate unstructured 3-dimensional data, see][]{Python}.

\bigskip
\noindent To investigate the degeneracies of this technique we considered a 0.5, 0.7, and 1.0 $\rm M_{\odot}$ mass star and a 2 and 5 $\rm M_{\rm jup}$ mass planet and ran a grid constituted of $\rm P_{rot,init}$= 1-11 days (with $\Delta \rm P_{rot,init}$ = 0.2 days) and $\rm SMA_{init}$ between 0.1 and 1.0  $\rm R_{co}$ \citep[see][]{Gallet18}. Figure \ref{Degemass3D} shows $\rm P_{\rm rot,\star}$(t) as a function of SMA(t), $\rm M_{\star}$, and $\rm M_p$, and displayed the age as a color gradient. It shows that the solutions are degenerated when using only $\rm P_{\rm rot,\star}$ or SMA, but that these degeneracies are lifted when using both quantities simultaneously. Additionally, the information about the fact that the planet is still (or not) orbiting the star adds another criteria and helps the lift of these degeneracies. 


In the case of migrating planets, the evolution of star-planet systems in the $\rm P_{\rm rot,\star}$(t)-SMA(t) space depends on their initial conditions. As the star evolves, its surface rotation rate is impacted by the inward migration of the planet. The star-planet system thus moves toward smaller SMA and rotation period (if the acceleration torque produced by the planet is stronger than the breaking torque of the stellar winds). 


%
%
%
%

\bigskip
\noindent We applied our technique on the planetary system WAPS-43, a 0.7 $\rm M_{\odot}$ K7V stars around which orbits a 2 $\rm M_{\rm jup}$ mass planet \citep{Hellier11}. The rotation period of the star is estimated at 15.6 $\pm$ 0.4 days and the planet is observed at a distance of 0.01526 au from the star. 

For this system we expect to get a small range of possible age and the fact that the planet is observed indicates that no engulfment has occurred, which discard the models in which the planet is engulfed by the star. 

Using the gyrochronology technique the age of WASP-43 is estimated at 400 Myr \citep{Hellier11}. The most probable solutions of our technique aim towards a more advanced system with an age between 3 and 6 Gyr. A 3D linear interpolation of the observed couple leads to an age estimation of $4.9^{+0.3}_{-0.4}$\,Gyr, which is consistent with our $S^2$ exploration. {With this technique we can also constrain the initial condition of the WASP-43 system. According to our simulations the initial rotation rate of the star was between 8 and 11 days (the slow part of the initial observed rotational distribution) and the planet was initially located between 0.025 and 0.03 au. }

\section{Conclusions}

In massive close-in planetary systems, the tidal interaction between the central star and the planet is expected to have strongly modified the evolution of the surface rotation rate of the host star. In that case, it is no longer possible to use age determination techniques based on stellar surface rotation and magnetic field. The corresponding forbidden region is located quite close from the stellar surface (few hundredth of au) and tends to expend for increasing stellar and planetary mass. 

To overcome this issue, we proposed a new age determination techniques that can be applied to such massive close-in planetary systems: the tidal-chronology. {This technique is based on the uniqueness of the path followed by a planetary system on the $\rm P_{\rm rot,\star}$(t)-SMA(t) plane.}

However, the numerical and physical description of the tidal dissipation in stellar and planetary interior is currently not good enough to only use this age estimation, which should be considered with caution and in combination with other age determination techniques (as any empirical age determination techniques). {Indeed, in this work we do not included the dissipation in the planet (that is still badly theoretically constrained) and the magnetic star-planet interaction \citep[see][]{Strugarek17}. The dissipation inside of the stellar radiative core is also currently not physically described and hence not numerically included. Nevertheless, we developed a promising technique that will benefit the community when all aspect of tidal and magnetic star-planet interactions will be included in angular momentum evolution models.}

These dissipations could produce an additional torque that could increase the rate of the planetary migration, and consequently reduce the estimated age of an observed planetary system. The impact of the planetary companion on the rotational evolution of the host stars should increase with the inclusion of the neglected effects mentioned above. Hence, the take-home message of our work is thus to be careful and aware of the limits of the gyrochronology analysis when using it to estimate the age of planetary systems.



\begin{acknowledgements}
We thank the SF2A 2018 organizers as well as Roxanne Ligi, Yveline Lebreton, Tristan Guillot, and Magali Deleuil for the organisation of the S17 session. F.G acknowledges financial support from the CNES fellowship. 
\end{acknowledgements}

\bibliographystyle{aa}  
\bibliography{references} 

\newcommand{\noop}[1]{}
\begin{thebibliography}{18}
\expandafter\ifx\csname natexlab\endcsname\relax\def\natexlab#1{#1}\fi

\bibitem[{{Alibert} {et~al.}(2013){Alibert}, {Carron}, {Fortier}, {Pfyffer},
  {Benz}, {Mordasini}, \& {Swoboda}}]{Alibert13}
{Alibert}, Y., {Carron}, F., {Fortier}, A., {et~al.} 2013, \aap, 558, A109

\bibitem[{{Amard} {et~al.}(2016){Amard}, {Palacios}, {Charbonnel}, {Gallet}, \&
  {Bouvier}}]{Amard15}
{Amard}, L., {Palacios}, A., {Charbonnel}, C., {Gallet}, F., \& {Bouvier}, J.
  2016, \aap, 587, A105

\bibitem[{{Barnes}(2003)}]{Barnes03}
{Barnes}, S.~A. 2003, \apj, 586, 464

\bibitem[{{Bolmont} \& {Mathis}(2016)}]{Bolmont16}
{Bolmont}, E. \& {Mathis}, S. 2016, Celestial Mechanics and Dynamical
  Astronomy, 126, 275

\bibitem[{{Delorme} {et~al.}(2011){Delorme}, {Collier Cameron}, {Hebb},
  {Rostron}, {Lister}, {Norton}, {Pollacco}, \& {West}}]{Delorme11}
{Delorme}, P., {Collier Cameron}, A., {Hebb}, L., {et~al.} 2011, \mnras, 413,
  2218

\bibitem[{{Gallet} {et~al.}(2018){Gallet}, {Bolmont}, {Bouvier}, {Mathis}, \&
  {Charbonnel}}]{Gallet18}
{Gallet}, F., {Bolmont}, E., {Bouvier}, J., {Mathis}, S., \& {Charbonnel}, C.
  2018, ArXiv e-prints

\bibitem[{{Gallet} {et~al.}(2017{\natexlab{a}}){Gallet}, {Bolmont}, {Mathis},
  {Charbonnel}, \& {Amard}}]{Gallet17b}
{Gallet}, F., {Bolmont}, E., {Mathis}, S., {Charbonnel}, C., \& {Amard}, L.
  2017{\natexlab{a}}, \aap, 604, A112

\bibitem[{{Gallet} \& {Bouvier}(2015)}]{GB15}
{Gallet}, F. \& {Bouvier}, J. 2015, \aap, 577, A98

\bibitem[{{Gallet} {et~al.}(2017{\natexlab{b}}){Gallet}, {Charbonnel}, {Amard},
  {Brun}, {Palacios}, \& {Mathis}}]{Gallet17}
{Gallet}, F., {Charbonnel}, C., {Amard}, L., {et~al.} 2017{\natexlab{b}}, \aap,
  597, A14

\bibitem[{{Hellier} {et~al.}(2011){Hellier}, {Anderson}, {Collier Cameron},
  {Gillon}, {Jehin}, {Lendl}, {Maxted}, {Pepe}, {Pollacco}, {Queloz},
  {S{\'e}gransan}, {Smalley}, {Smith}, {Southworth}, {Triaud}, {Udry}, \&
  {West}}]{Hellier11}
{Hellier}, C., {Anderson}, D.~R., {Collier Cameron}, A., {et~al.} 2011, \aap,
  535, L7

\bibitem[{{Ida} \& {Lin}(2008)}]{Ida08}
{Ida}, S. \& {Lin}, D.~N.~C. 2008, \apj, 685, 584

\bibitem[{Jones {et~al.}(2001)Jones, Oliphant, Peterson, {et~al.}}]{Python}
Jones, E., Oliphant, T., Peterson, P., {et~al.} 2001, {SciPy}: Open source
  scientific tools for {Python}

\bibitem[{{Lanza} {et~al.}(2011){Lanza}, {Damiani}, \& {Gandolfi}}]{Lanza11}
{Lanza}, A.~F., {Damiani}, C., \& {Gandolfi}, D. 2011, \aap, 529, A50

\bibitem[{{Mordasini} {et~al.}(2012){Mordasini}, {Alibert}, {Benz}, {Klahr}, \&
  {Henning}}]{Mordasini12}
{Mordasini}, C., {Alibert}, Y., {Benz}, W., {Klahr}, H., \& {Henning}, T. 2012,
  \aap, 541, A97

\bibitem[{{Mordasini} {et~al.}(2009){Mordasini}, {Alibert}, {Benz}, \&
  {Naef}}]{Mordasini09}
{Mordasini}, C., {Alibert}, Y., {Benz}, W., \& {Naef}, D. 2009, \aap, 501, 1161

\bibitem[{{Skumanich}(1972)}]{Sku72}
{Skumanich}, A. 1972, \apj, 171, 565

\bibitem[{{Strugarek} {et~al.}(2017){Strugarek}, {Bolmont}, {Mathis}, {Brun},
  {R{\'e}ville}, {Gallet}, \& {Charbonnel}}]{Strugarek17}
{Strugarek}, A., {Bolmont}, E., {Mathis}, S., {et~al.} 2017, \apjl, 847, L16

\bibitem[{{Vidotto} {et~al.}(2014){Vidotto}, {Gregory}, {Jardine}, {Donati},
  {Petit}, {Morin}, {Folsom}, {Bouvier}, {Cameron}, {Hussain}, {Marsden},
  {Waite}, {Fares}, {Jeffers}, \& {do Nascimento}}]{Vidotto14}
{Vidotto}, A.~A., {Gregory}, S.~G., {Jardine}, M., {et~al.} 2014, \mnras, 441,
  2361

\end{thebibliography}

\end{document}